# The 2011 unrest at Katla volcano: characterization and interpretation of the tremor sources

Giulia Sgattoni[1,2,3*], Ólafur Gudmundsson[3], Páll Einarsson[2], Federico Lucchi[1], Ka Lok Li[3], Hamzeh Sadeghisorkhani[3], Roland Roberts[3], Ari Tryggvason[3]

[1] *Department of Biological, Geological and Environmental Sciences, University of Bologna, Bologna, Italy*
[2] *Institute of Earth Sciences, Science Institute, University of Iceland, Reykjavik, Iceland*
[3] *Department of Earth Sciences, Uppsala University, Uppsala, Sweden*

*Corresponding author: giulia.sgattoni2@unibo.it*

## Abstract

A 23 hour tremor burst was recorded on July 8-9th 2011 at the Katla subglacial volcano, one of the most active and hazardous volcanoes in Iceland. This was associated with deepening of cauldrons on the ice cap and a glacial flood that caused damage to infrastructure. Increased earthquake activity within the caldera started a few days before and lasted for months afterwards and new seismic activity started on the south flank. No visible eruption broke the ice and the question arose as to whether this episode relates to a minor subglacial eruption with the tremor being generated by volcanic processes, or by the flood. The tremor signal consisted of bursts with varying amplitude and duration. We have identified and described three different tremor phases, based on amplitude and frequency features. A tremor phase associated with the flood was recorded only at stations closest to the river that flooded, correlating in time with rising water level observed at gauging stations. Using back-projection of double cross-correlations, two other phases have been located near the active ice cauldrons and are interpreted to be caused by volcanic or hydrothermal processes. The greatly increased seismicity and evidence of rapid melting of the glacier may be explained by a minor sub-glacial eruption. It is also plausible that the tremor was generated by hydrothermal boiling and/or explosions with no magma involved. This may have been induced by pressure drop triggered by the release of water when the glacial flood started. All interpretations require an increase of heat released by the volcano.

**Keywords:** Katla volcano, unrest, volcanic tremor, flood tremor, sub-glacial eruption, hydrothermal boiling

# 1. Introduction

A wide range of seismic signals are recorded at volcanoes, generated by several processes, including magmatic, geothermal or tectonic processes. Volcanic tremor and long-period (LP, Chouet, 1996) events are thought to be generated by fluid movements in association with volcanic eruptions or hydrothermal activity (Chouet, 2003; McNutt, 2005). Since they often precede or accompany eruptions, understanding these phenomena is of major interest for volcano monitoring. The precise mechanisms responsible for generating those signals are, however, still debated, particularly for tremor, which is a more complex signal compared to earthquakes and is difficult to locate as picking first arrivals is usually not possible (Konstantinou and Schlindwein, 2002).

Volcanic tremor is a persistent seismic signal, observed only at active volcanoes, that lasts several minutes to several days and is observed prior to or during most volcanic eruptions (Fehler, 1983; Julian, 1994; Ripepe, 1996; Métaxian et al., 1997). Because of the absence of clear onsets of waves, volcanic tremor cannot be located with conventional arrival time methods. Therefore, other techniques have been developed, based for example on phase coherency of signals among stations, either with seismic arrays (e.g. Furumoto et al., 1990; Goldstein and Chouet, 1994; Métaxian et al., 1997) or with sparse seismic networks (Guðmundsson and Brandsdóttir, 2010; Ballmer et al., 2013; Droznin et al., 2015; Li et al., submitted for publication). Alternatively, the signal amplitude distribution at different stations can be used to infer the source location (e.g. Battaglia and Aki, 2003; Di Grazia et al., 2006), when the amplitude decay with distance has a simple pattern.

A variety of source models have been proposed for the generation of volcanic tremor, including excitation and resonance of fluid-filled cracks (Chouet, 1992; Benoit and McNutt, 1997), fluid-flow-induced oscillations of conduits transporting magmatic fluids (Julian, 1994), magma column wagging (Jellinek and Bercovici, 2011) and frictional flow (Dmitrieva et al., 2013). There are also observations of tremor composed by interfering, closely-spaced LP events, e.g. at Montserrat (Baptie et al., 2002). Tremor resembling volcanic tremor has been recorded in association with hydrothermal activity, defined as 'non-eruption tremor' by Leet (1988). This is generated by bubble growth or collapse due to hydrothermal boiling. A similar source has been suggested for geothermal noise at Ölkelduháls, SW Iceland (Guðmundsson and Brandsdóttir, 2010). Tremor has been recorded also in association with hydrothermal explosions occurring as a consequence of pressure drop on the hydrothermal system, triggered for example by landslides (Jolly et al., 2014) or a sudden level drop of a volcanic lake (Montanaro et al., 2016).

At subglacial volcanoes, a number of other processes related to glacier dynamics can be responsible for producing seismic signals. In addition to glacial earthquakes, due to e.g. glacier sliding, crevassing, ice falls (e.g. Métaxian et al., 2003; Jónsdóttir et al., 2009), seismic tremor can be generated in association with glacial floods, which in turn can be caused by subglacial volcanic eruptions or by drainage of subglacial lakes formed by geothermal activity (Guðmundsson et al., 2008). Seismic tremor generated by jökulhlaups (glacial floods) has been recorded several times in Iceland, e.g. from the Skaftá cauldrons on Vatnajökull ice cap (Zóphóníasson and Pálsson, 1996) during the Gjálp eruption in 1996 (Einarsson et al., 1997), but no detailed description is available in the literature.

Sometimes, glacial and volcanic processes act together and it is difficult to discern whether or not an eruption has started, or even occurred at all, under a glacier. It is very important for volcano monitoring to be able to discern between these different signals and processes, particularly in Iceland, where a number of large volcanoes are covered by ice. One of these is Katla, located under Mýrdalsjökull glacier in south Iceland, just east of Eyjafjallajökull. After the eruption of Eyjafjallajökull in 2010, attention was attracted to the neighbouring Katla, as some previous eruptions of Eyjafjallajökull had been followed by more powerful eruptions of Katla (Einarsson and Hjartardóttir, 2015). The two volcanoes are tectonically connected (Einarsson and Brandsdóttir, 2000) and are both covered by glaciers. Therefore, their volcanic activity is dominated by explosive eruptions due to magma - ice interaction. The erupted volumes from Katla have been at least one order of magnitude bigger than its neighbour's (Sturkell et al., 2010).

Katla's last eruption to break the ice surface occurred in 1918 and the present repose time is the longest known in historical times (Larsen, 2000). However, two minor sub-glacial eruptions with no tephra emission into the atmosphere possibly occurred in 1955 and 1999. Both were accompanied by formation of new ice depressions (cauldrons) on the ice surface and jökulhlaups, but no visible eruption through the ice. During the 1999 unrest, a tremor signal was recorded (e.g. Thorarinsson, 1975; Sigurðsson et al., 2000). A similar episode occurred recently in July 2011, when increased earthquake activity was recorded, together with a tremor burst and a jökulhlaup (glacial flood) that drained from south-east Mýrdalsjökull and destroyed a bridge on the main road. One important question is whether this unrest is due to a minor subglacial eruption and the tremor associated to volcanic processes (magmatic or hydrothermal) or to the flood.

In this article we analyse the 2011 tremor signal and associated earthquake activity. We use cross-correlation methods to extract information about the source location. We study the time evolution of tremor attributes, such as amplitude and frequency, and compare them with direct hydrological observations and the evolution of the earthquake activity, to constrain possible sources.

## 2. The Katla volcanic system

The Katla volcanic system consists of a central volcano with a 110 km² summit caldera (up to 14-km wide; Fig. 1) filled with the 600 to 750 m thick ice of Mýrdalsjökull glacier (Björnsson et al., 2000) and the Eldgjá fissure system which extends 75 km to the northeast (Larsen, 2000; Thordarson et al., 2001; Fig. 1). It is located south of the intersection between the Eastern Volcanic Zone and the transform boundary of the South Iceland Seismic Zone (Sturkell et al., 2008) and forms a part of the Eastern Volcanic Zone.

The caldera wall is breached in three places, to the south-east, north-west and south-west. These gaps provide outflow paths for ice in the caldera to feed the main outflow glaciers, Kötlujökull, Entujökull and Sólheimajökull and are the potential pathways for melt water from the glacier in jökulhlaups (Sturkell et al., 2010). Several ice cauldrons (at least 16) are located within and at the caldera rim, representing the surface expression of subglacial geothermal activity. Changes in their geometry are monitored to detect variations of geothermal heat release (Guðmundsson et al., 2007).

A velocity anomaly at shallow depth, revealed by seismic undershooting, was interpreted as evidence of a magma chamber (Guðmundsson et al., 1994) and a non-magnetic body was identified within the same region with an aeromagnetic survey (Jónsson and Kristjánsson, 2000). The presence of a magma chamber is supported by geobarometry analyses on historical tephra samples, conducted by Budd et al. (2014), but is questioned by tephra stratigraphy studies by Óladóttir et al. (2008).

**2.1. Recent activity and seismicity**

Apart for the large Eldgjá lava eruption in AD 934-940 (Thordarson et al., 2001), all historical eruptions of the Katla volcanic system, at least 20, occurred within the caldera (Larsen, 2000) and consisted mainly of basaltic phreatomagmatic eruptions, capable of producing destructive glacial floods. The last eruption to break the ice-surface was an explosive basaltic eruption in 1918. It lasted for about three weeks and was accompanied by a massive jökulhlaup (Sveinsson, 1919). The eruption site was located near the southern rim of the caldera, beneath about 400 m of ice. The height of the eruptive plume was estimated as 14 km a.s.l. (Eggertsson, 1919), the volume of tephra fall-out around 0.7 km$^3$ and the volume of water-transported material is estimated to be between 0.7 and 1.6 km$^3$ (Larsen, 2000).

Persistent seismic activity has been observed at Katla since the first sensitive seismographs were installed (in the 1960s). When reasonably accurate locations became available, the activity was shown to occur in two distinct main areas: within the caldera and at Goðabunga on the western flank (Einarsson and Brandsdóttir, 2000). The seismicity inside the caldera consists of high frequency and hybrid events, probably associated with the subglacial geothermal activity (Sturkell et al., 2010) and volcano-tectonic processes. The Goðabunga cluster consists mainly of long-period shallow events and has a controversial interpretation, as a response to a slowly-rising viscous crypto-dome (Soosalu et al., 2006) or in association with ice fall events (Jónsdóttir et al., 2009). Periods of high seismicity were observed in 1967 and 1976-77 (Einarsson, 1991). In the 1976-77 episode, both epicentral areas were active, i.e. both within the caldera and at Goðabunga.

Two possible minor sub-glacial eruptions occurred in 1955 and 1999, both at the caldera rim. The 1955 event took place at the eastern rim of the caldera. Two shallow ice-cauldrons formed and a small jökulhlaup drained from Kötlujökull, i.e. south-east Mýrdalsjökull (Thorarinsson, 1975). The 1999 event took place in July at the south-western rim. Seismic stations around the glacier recorded earthquakes and bursts of tremor that culminated in the release of a jökulhlaup from Sólheimajökull, i.e. south-west

Mýrdalsjökull (Sigurðsson et al., 2000; Roberts et al., 2003). A new cauldron also formed on the surface of the glacier (Guðmundsson et al., 2007).

From 1999 to 2004, uplift of the volcano was revealed by GPS measurements on the caldera rim, interpreted to result from 0.01 km³ magma accumulation (Sturkell et al., 2006; 2008). This interpretation was supported by the evidence of increased geothermal heat output observed in 2001-2003, based on the evolution of ice cauldrons (Guðmundsson et al. (2007). A recent study by Spaans et al. (2015), suggested, instead, that the 1999-2004 uplift may be due to glacial isostatic adjustment as a consequence of mass loss of Iceland's ice caps.

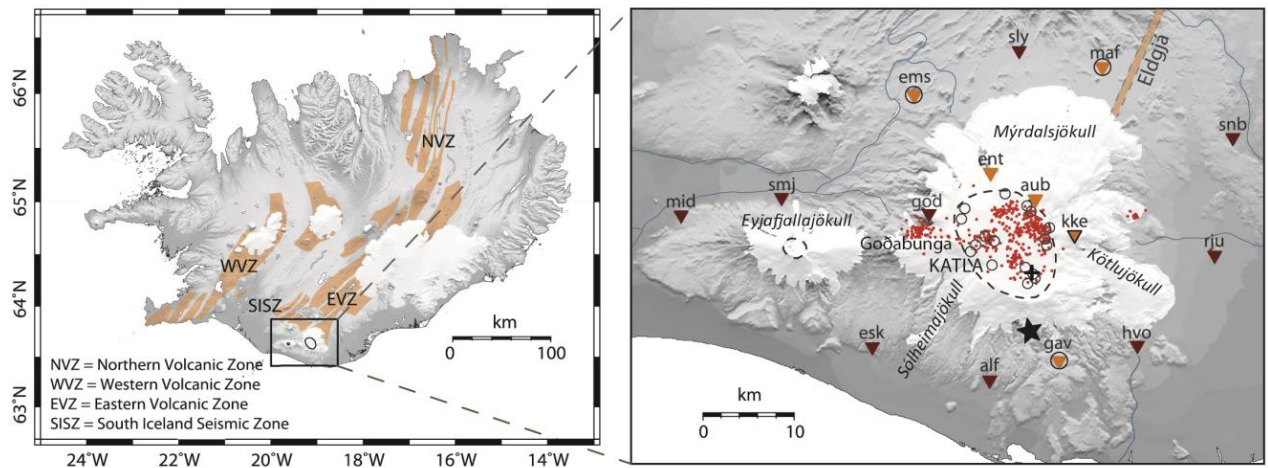

Fig. 1. Map of Iceland showing the different volcanic systems (in orange; from Einarsson and Sæmundsson, 1987). In the inset, the seismic network operating during the tremor and the main seismic and geological features of Katla are shown. Dark brown triangles: permanent Icelandic Meteorological Office (IMO) seismic stations. Orange triangles: temporary Uppsala University seismic stations operating during 29 May 2011 - August 2013 (no outline), during 6 July 2011 - August 2014 (black outline) and during 10 July 2011 – August 2013 (black circle). Red dots: epicentres at Katla before July 2011. These are mostly localized in two distinct source areas, within the caldera and on the west flank at Goðabunga. The star marks the new cluster on the south flank, started in July 2011 (Sgattoni et al., 2016). The Katla and Eyjafjallajökull caldera rims are outlined by dashed lines. Open circles correspond to ice cauldrons in the Mýrdalsjökull glacier (Guðmundsson et al., 2007). White areas are glaciers. To the NE, the location of Eldgjá fissure is shown. The black cross in the south-eastern caldera marks the site of the 1918 eruption.

## 3. July 2011 unrest: Course of events

Between August 2010 and July 2011, most of the ice cauldrons on the Mýrdalsjökull glacier uplifted by 6 to 8 m, as interpreted by Guðmundsson and Sólnes (2013) due to water accumulation under the glacier. The greatest rise, 11-12 m was observed at cauldron 16 (Fig. 2; Guðmundsson and Sólnes, 2013).

Since the beginning of July 2011 the seismicity intensified, especially within the caldera, and culminated on July 8-9th, when continuous tremor was recorded, starting at ~19:00 GMT on July 8th. No signs of eruption breaking the ice were seen, but a jökulhlaup drained from Kötlujökull and deepening of some ice cauldrons was observed on the surface of Mýrdalsjökull in the southern and eastern parts of the caldera (Fig. 2). The jökulhlaup (~18 million m3) swept away the bridge on the main road 1 over Múlakvísl river around 05:00 GMT on July 9th, one hour after rising water level was detected at the gauging station Léreftshöfuð, located a few km south of Kötlujökull (IMO, 2011) and ~6 km upstream from the bridge.

Another gauging station, located on the bridge over Múlakvísl river, began to show slightly increased conductivity, reaching values close to 200 $\mu S/cm$, around midnight on July 7th. This doesn't coincide with increased water level. The conductivity later rose again above 200 $\mu S/cm$ at around midnight on July 8th and a dramatic increase occurred after midnight on July 9th, around the time of maximum of the tremor (IMO, 2011), coinciding with dramatic water level increase (Fig. 2).

Galeczka et al., (2014) conducted a study of the chemical composition of the flood water and did not find evidence that the water had come into contact with magma. They suggested, therefore, that the heat source for glacier melting was geothermal rather than magmatic.

In association with the tremor, a new source of seismic events was activated on the south flank of Katla, at the southern edge of Mýrdalsjökull glacier (Fig. 2). This seismicity has been interpreted as related to a new hydrothermal system that may have activated on Katla's south flank during this unrest episode, although no new hydrothermal area was found (Sgattoni et al., 2016; Sgattoni et al., submitted for publication). The course of events occurred in association with the 2011 unrest is reported in Table 1.

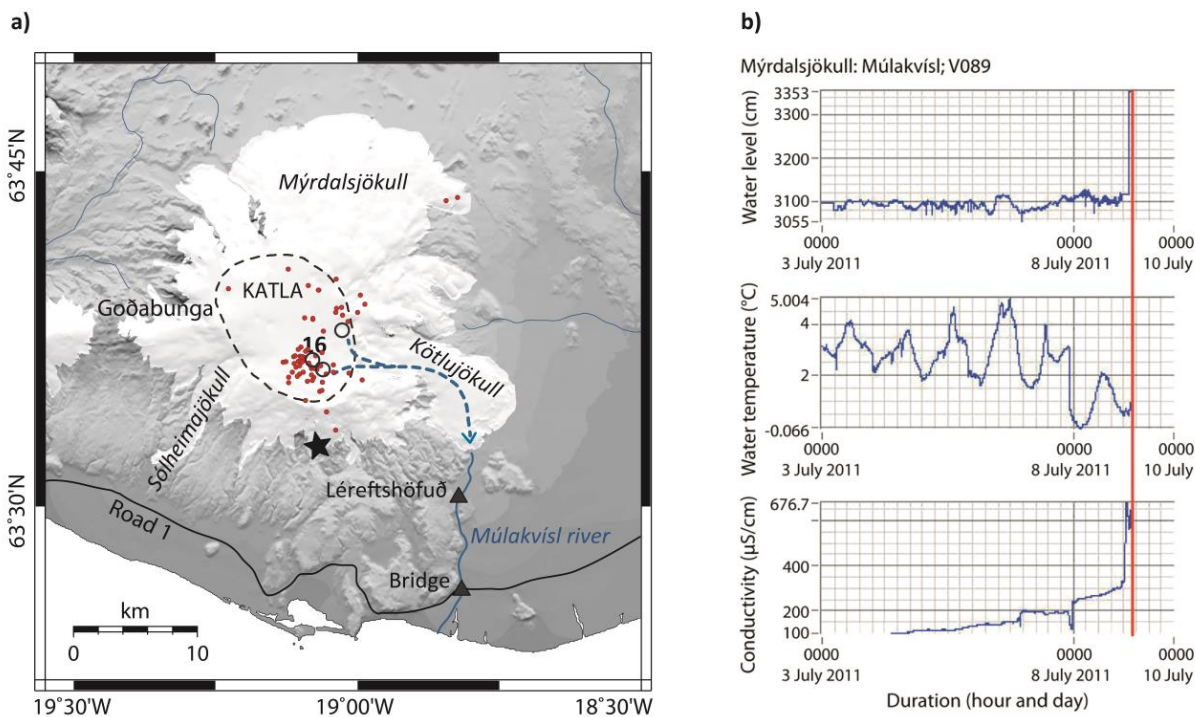

Fig. 2. a) Map of Katla showing the features related to the July 2011 unrest. Black open circles are the ice cauldrons that deepened during the unrest, number 16 is the one that showed the biggest change before and during the unrest. The dashed blue arrow shows the presumed flood path. Red dots are the earthquakes that occurred on July 8th and 9th. The 2 gauging stations are marked with black triangles, the southern one corresponding to the bridge over Múlakvísl river. The star marks the new seismic cluster on the south flank (Sgattoni et al., 2016). b) IMO continuous monitoring of water level, temperature and conductivity of Múlakvísl river. Data from the gauging station located at the bridge over the river. The red line marks the time when the station stopped working due to the flood that destroyed the bridge (IMO, 2015).

| **Before tremor** |
|---|
| - from August 2010: ice cauldrons uplifting |
| - from beginning of July: increased seismicity inside caldera and new seismicity on south flank |
| **8th – 9th July 2011** |
| - July 8th, 00:00: conductivity above 200 µS/cm in Múlakvísl river |
| - July 8th, ~19:15: tremor starts |
| - July 9th, 00:00: dramatic increase in conductivity in Múlakvísl river |
| - July 9th, 04:00: rising water level detected at gauging station Léreftshöfuð |
| - July 9th, 05:00: jökulhlaup destroys a bridge on road n.1 |
| - July 9th, 18:00: tremor ends |

Table 1. Course of events that occurred in association with the July 2011 unrest at Katla.

## 4. Seismic network

Following the eruption of Eyjafjallajökull volcano in 2010, the seismic monitoring network run by the Icelandic Meteorological Office (IMO) around Katla was densified from 5 to 9 stations. Moreover, Uppsala University deployed additional 9 temporary stations between May-October 2011 and August 2013 (Fig 1). Three of these stations, located along the caldera rim, were deployed before the tremor episode occurred. The rest were deployed later and therefore could not be used to analyse the tremor. However, 3 of them were deployed immediately after the tremor episode, and could therefore be used to locate the earthquakes after July 9th.

Of the total 15 seismic stations used in this study, 8 stations were equipped with broadband Guralp ESPA and Guralp CMG3-ESPC sensors with a flat response from 60 s to the Nyquist frequency (50 Hz). The remaining 7 stations had 5-second Lennartz sensors. Data were recorded and digitized with Guralp and Reftek systems at 100 Hz. Stations were powered with batteries, wind generators and solar panels. All the instruments recorded in continuous mode, but some technical problems (e.g. power failure) mainly due to harsh weather condition (especially in winter time), prevented some stations from working continuously during the whole operation time. Data from a total of 10 seismic stations near the caldera and around the glacier were used to locate and analyse the tremor signal (Fig 1).

## 5. Earthquake activity

The seismic events within the caldera and at Goðabunga that occurred prior to, during and after the tremor, between June 20th and July 20th have been automatically detected and located with the SIL analysis software by using manually picked P- and S-phases (Böðvarsson et al., 1998). The software uses a single event location technique, performed by minimizing the square sum of both P- and S- wave residual arrival times in a 1D velocity model. The velocity model was obtained from tomographic studies of the area (Jeddi et al., 2015). Data for the seismic events located on the south flank, instead, were obtained with cross-correlation methods, described in Sgattoni et al. (2016). The number of seismic stations changed during the time period analysed from 11 to 12 stations on July 6th (when station KKE was installed) and from 12 to 15 stations immediately after the tremor.

The cumulative number of events detected is shown in Fig. 3, separately for the three main clusters identified at Katla: Inside the caldera, at Goðabunga and on the southern flank. A sharp increase of seismicity inside the caldera occurred on July 7-8th, together with the onset of the seismicity on the south flank. The Goðabunga cluster, instead, appears to have become less active. The increased seismicity cannot be explained only by the additional station deployed on July 6th, as can be seen in Fig. 4, where the seismicity clearly increased already on July 6th and maintained a high rate also after the tremor. A small swarm also occurred around 02:00 on July 6th (Fig. 4). After the tremor event, the seismicity remained high, but decreased slowly (Fig. 3).

We have located only events with at least 5 identified phases (P and S), for a total of 480 events between June 20th and July 20th, of which 56 occurred during the tremor (Fig. 5). However, many more earthquakes can be visually identified in the seismograms (more than 80 events during the tremor) but could not be located and are therefore not counted in the cumulative plot of Fig.3 because they were only recorded at the closest stations. This is clear when comparing the seismograms of Fig. 4: The intense seismicity at the caldera station AUB is not observed on the seismogram at station ALF, located several km south of the caldera (Fig. 4). At ALF station, however, the south-flank seismicity appears clearly, with LP events with a peculiar regular temporal pattern commencing on July 7th (Fig. 4). Only few similar, much smaller, events were observed in the months before, with no regular temporal pattern (Sgattoni et al., 2016).

The hypocentral locations of the 480 events are shown in Fig. 5. Most events are small in magnitude, with only ~100 events with magnitude M>1 (magnitudes from IMO catalogue). Based on the temporal changes of the seismicity, we defined three different time intervals: 1) 20 June – 6 July, before the increased seismicity started; 2) 6-9 July, from the increased seismicity to the end of the tremor; 3) 10-20 July, after the tremor. During the first time period, the seismicity was distributed evenly inside the caldera and at Goðabunga (Fig. 5). Some seismic events were also located near the northern active cauldron (Fig. 5). After the seismicity increased (second time interval) most of the hypocentres were clustered in the south-eastern part of the caldera (Fig. 5). More than 30 events with magnitude M>1 occurred between July 6-9th. The largest event had magnitude equal to 2.3. After the tremor, the seismicity was mainly concentrated in the northern sector of the caldera (Fig. 5). The formal uncertainty of the locations is of the order of 1 km in the horizontal components and several km in the vertical. The distribution in depth is therefore not well resolved, as most of the stations used for the location were located at some distance from the hypocentres.

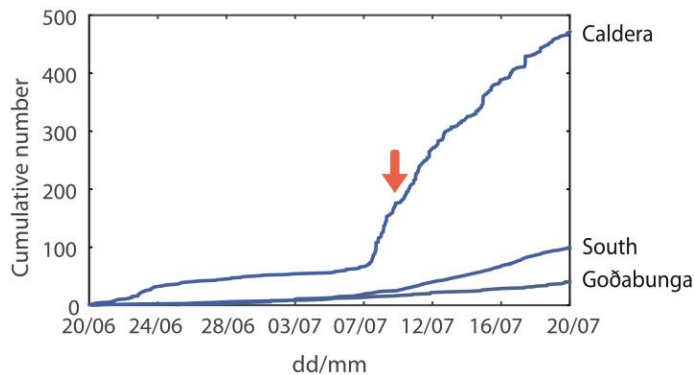

Fig. 3 Cumulative number of seismic events that occurred between June 20th and Jul 20th in the 3 main clusters at Katla. The arrow indicates the time of tremor. Data for the southern cluster are from Sgattoni et al. (2016).

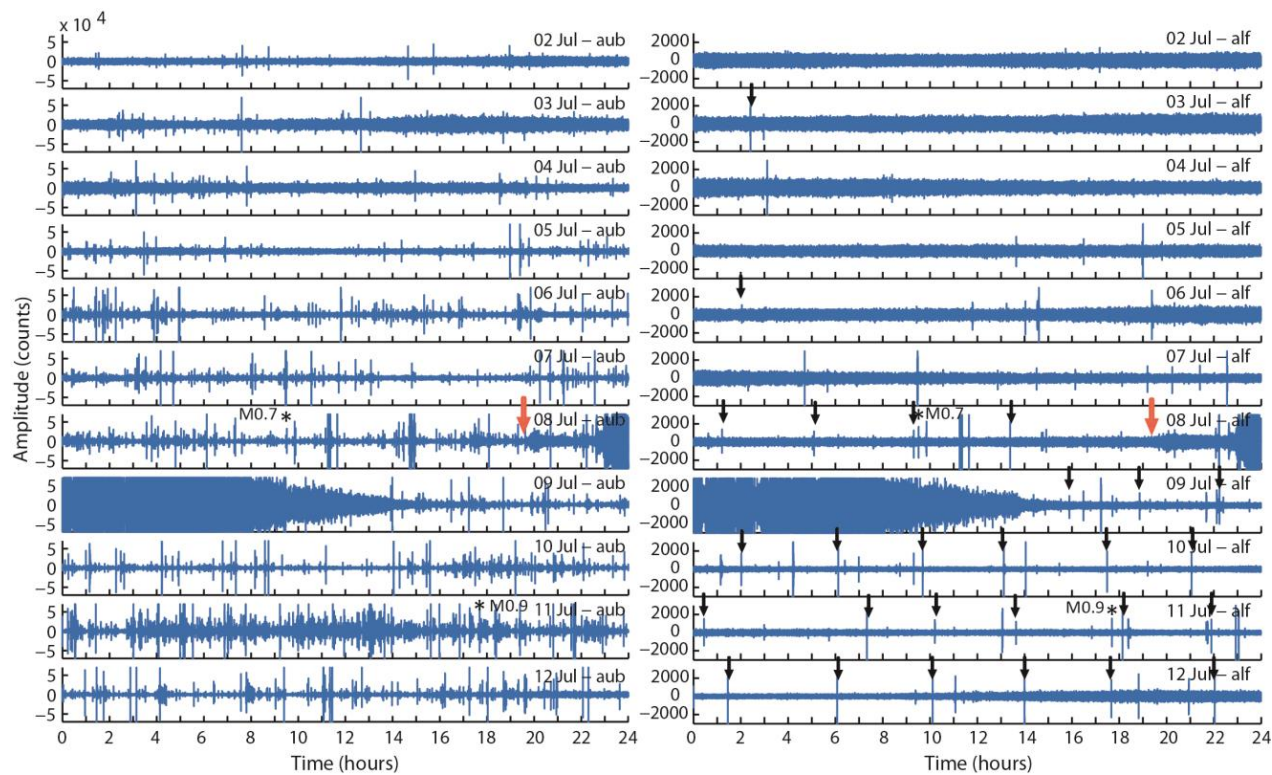

Fig. 4. Daily Z-component seismograms at stations AUB (left) and ALF (right) (see Fig. 1 for location) for the time period between July 2nd and July 12th. The orange arrow marks the onset of the tremor. The black arrows in the right panel mark the regular seismicity that started on the south flank of Katla (Sgattoni et al., 2016). The amplitude is in digital counts, proportional to velocity and the instrument response removed, so that amplitudes at the two stations are comparable. However the seismograms are clipped, for plotting reasons. The magnitudes of two events located in the south-eastern caldera are given as reference.

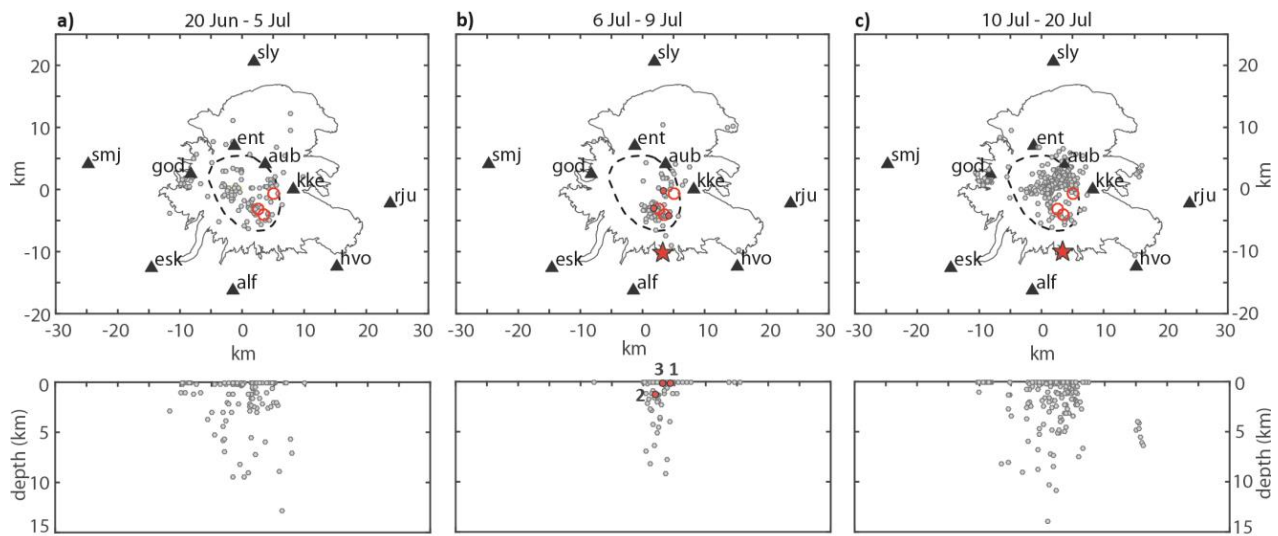

Fig. 5. Map location and depth distribution of events in three different time intervals: a) before the increased seismicity, b) after the increased seismicity until the end of the tremor, c) after the tremor. Red open circles: active cauldrons. Red star: new seismic cluster on the south flank. Red dots in panel b, labelled with 1,2,3: earthquakes shown in Figs. 6-8.

We will describe in more detail the waveforms of some events recorded after the seismicity picked up on 6-7th July. Most of the seismicity consisted of shallow events with varying frequency content. Most of the located events fall into two frequency ranges, 0.5-3 Hz and 0.5-10 Hz (same frequency as the tremor; see the next chapter). Before the tremor, the biggest events (high amplitude peaks in Fig. 4) fall in the second frequency range, while during the tremor there is a more even distribution of the two frequency ranges. In addition, several smaller high-frequency events, with frequency content up to 20 Hz, were recorded mainly at the closest stations (AUB and KKE). Many of the earthquakes during the intense seismicity period observed at AUB in Fig. 4 fall into this category.

Three events, among the biggest recorded during the tremor, are shown as examples in Figs. 6-8. Two of them were located in the south-eastern caldera, close to the southern cauldron that deepened during the unrest (Fig. 5). Of these, the first is characterised by low-frequency content with a main peak between 2-3 Hz, an emergent P wave and unclear S (Fig. 6). The second is composed of a wider spectrum of frequencies, up to 10 Hz, and can be classified as hybrid, with a high frequency beginning of the seismogram and lower frequency coda (Fig. 7). Noticeably, the higher frequencies appear to be strongly attenuated at the stations located on the other side of the caldera with respect to the source, particularly at the caldera stations KKE, AUB and ENT, the last being the most attenuated. The P wave is clear at ALF and GOD, emergent at KKE and AUB, unclear at ENT. The third event (Fig. 8), located near the northern active cauldron, contains frequencies mainly between 1-4 Hz, and has an emergent P wave and unclear S.

A conspicuous number of events was detected at station HVO after the river flooded, mainly consisting of small events with frequency content between 1 and 7 Hz. Further details about this seismicity are not reported as that is beyond the scope of this article.

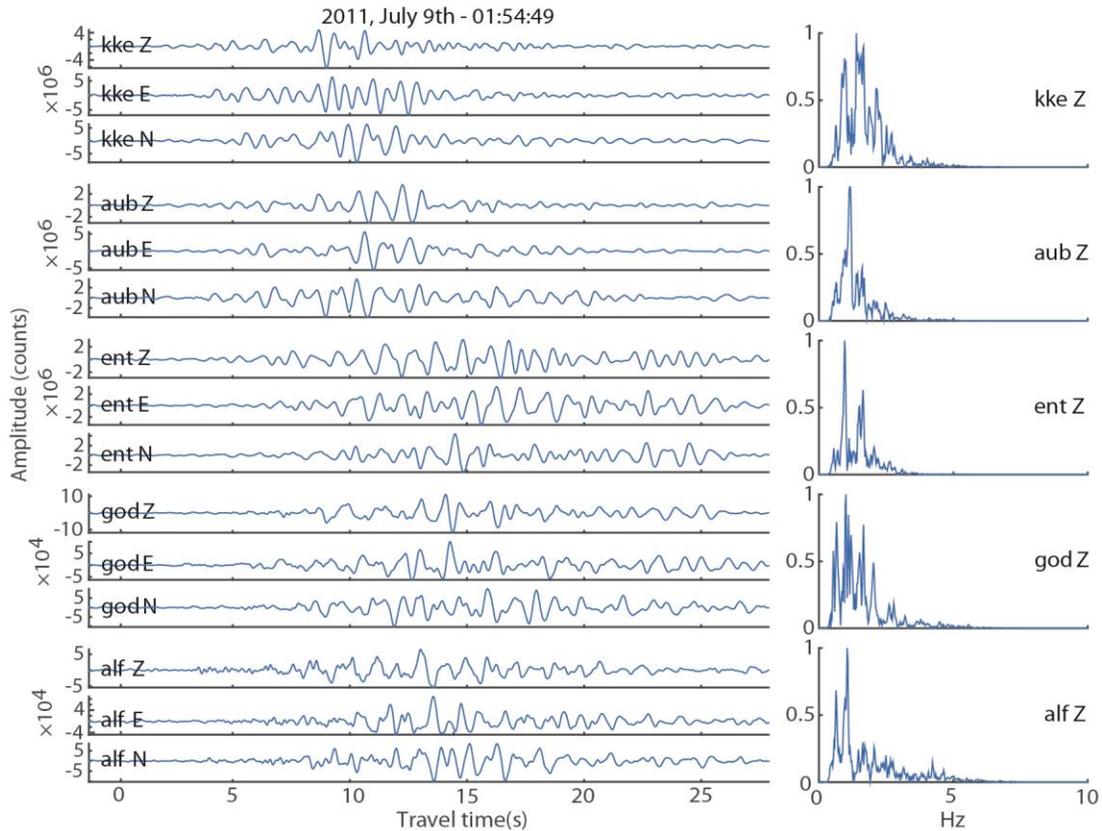

Fig. 6. Three component waveforms and Z component spectra of seismic event labelled '1' in Fig. 5. The amplitude is in digital counts, proportional to velocity. The magnitude is 1.56.

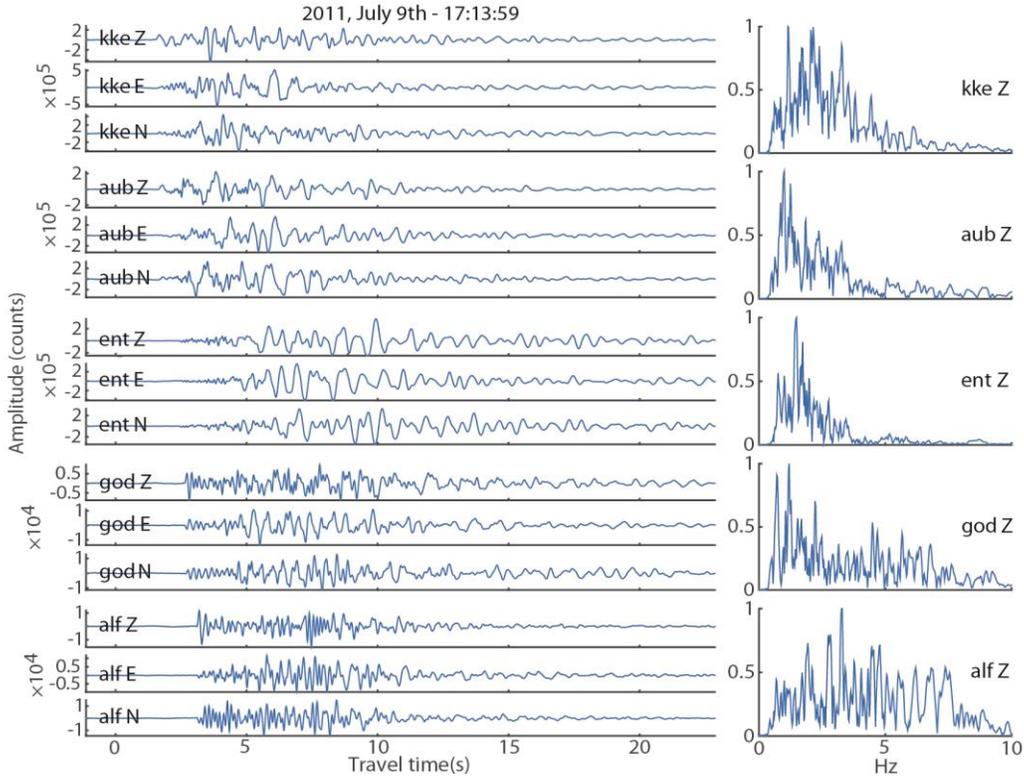

Fig. 7. Three component waveforms and Z component spectra of seismic event labelled '2' in Fig. 5. The amplitude is in digital counts, proportional to velocity. The magnitude is 1.61.

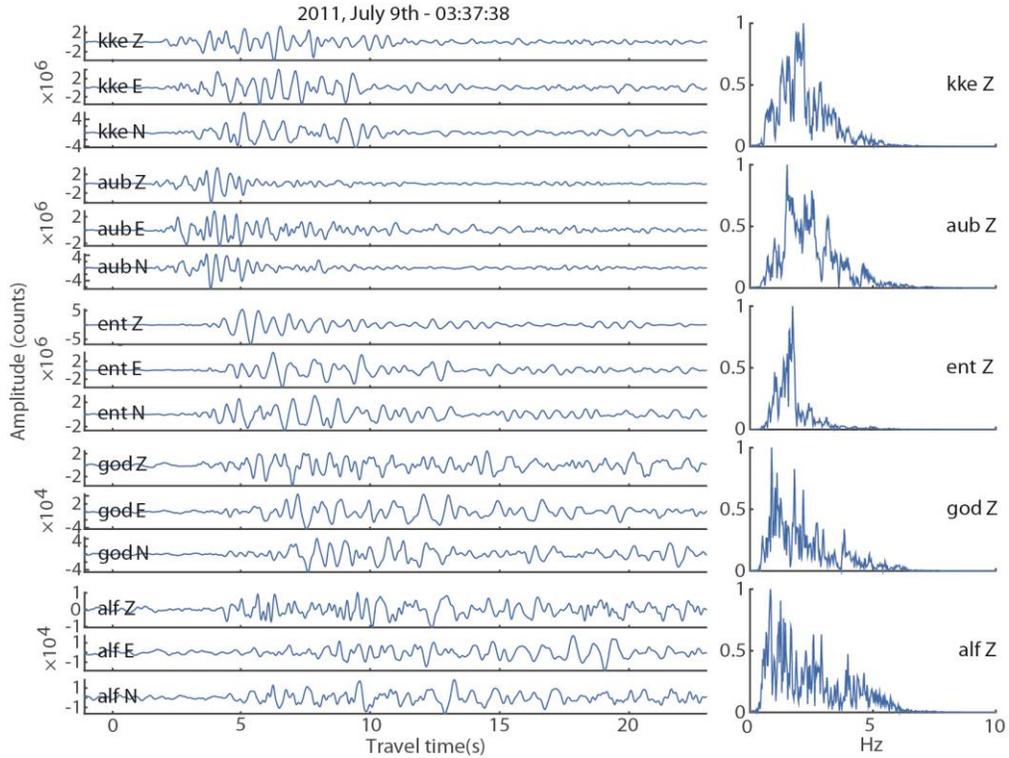

Fig. 8. Three component waveforms and Z component spectra of seismic event labelled '3' in Fig. 5. The amplitude is in digital counts, proportional to velocity. The magnitude is 1.65.

# 6. Frequency and amplitude characteristics of the tremor signal

## 6.1 Tremor pre-processing

The tremor signal is complex, with energy and frequency content varying with time. In order to study the tremor amplitude and frequency features, we first had to clean the signal from transients. As is evident from Fig. 4, persistent seismicity occurred during the days of the unrest, also during the tremor. Since many events have frequency content similar to the tremor, filtering is not effective. In addition, the amplitude time-features of the tremor change over time scales sometimes shorter than minutes, making the use of clipping strategies problematic. Therefore, in order to clean the signal from earthquakes, we used a combination of manual removal of earthquakes and clipping.

The signal has been examined in detail in order to identify local events, based on the time behaviour and on the frequency content. Seismic events usually manifest themselves in the signal as short transients with abrupt amplitude change compared to the 'background tremor'. They are also evident in the spectrograms, where they appear as peaks creating vertical lines with high amplitude compared to the adjacent time windows. By looking at both the raw seismograms and the spectrograms, we iteratively removed signals that resembled earthquakes. The time windows containing events were cut out of the signal and tapering was applied to the sides of the window. This was then accounted for when computing amplitude spectra, by normalizing the spectral amplitude based on the length of the windows that were cut out. Furthermore, we clipped the tremorgrams in order to suppress any leftover events. This process considerably reduced the number of sharp peaks in the spectrograms and in the amplitude time-history of the tremor.

## 6.2 Frequency content

The main features of the tremor signal can be seen in the spectrograms of Fig. 9, The tremor started around 19:00 GMT on July 8th and lasted for about 23 hours. The energy is distributed between 0.8-10 Hz, but mainly concentrated between 0.8-4 Hz, at most stations. The tremor was strongest between 23:00 GMT (July 8th) and 05:00 GMT (July 9th), when a number of short bursts occurred, ranging between 6 and 50 minutes in duration. Moreover, a distinct tremor phase with a broader frequency content is clear on the spectrogram at station HVO on July 9th. The time of this phase correlates with the surface water flooding and this station is located very close to the river that flooded.

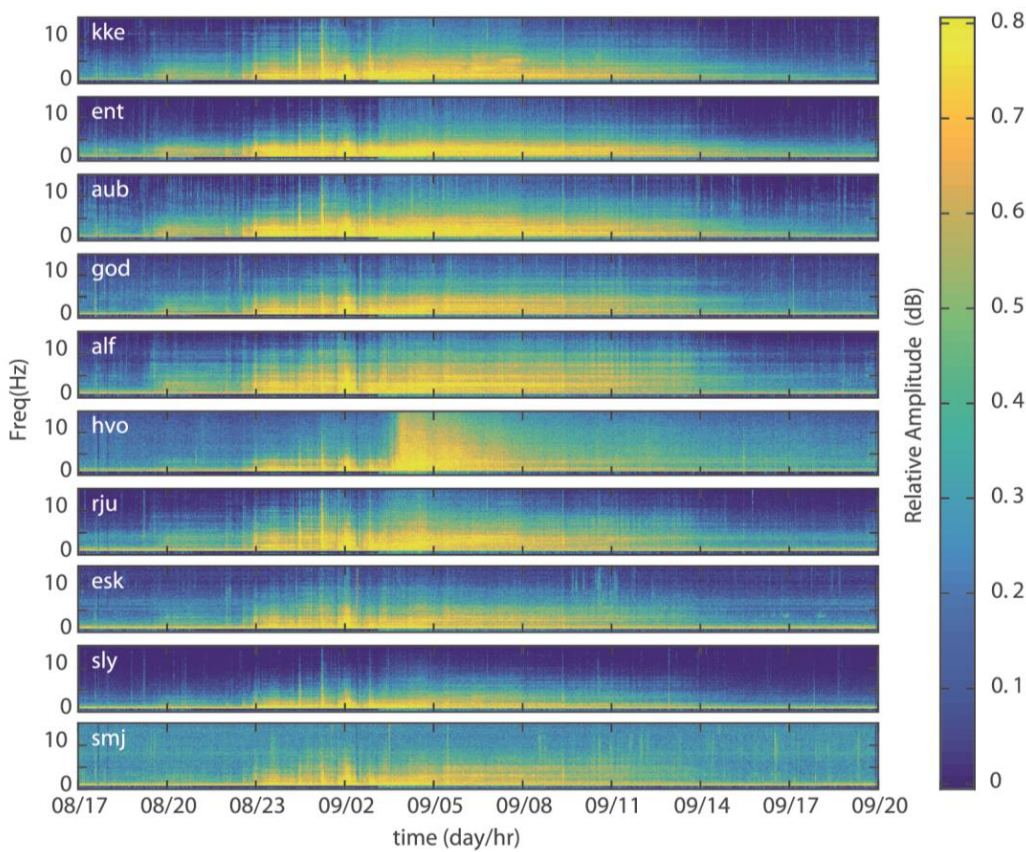

Fig. 9. Spectrograms of the Z component of motion at all stations. The signal was pre-filtered between 0.5-15 Hz. The colour scale shows the logarithmic relative amplitude (dB) individually normalised by maximum at each station. The stations are ordered from top to bottom by decreasing values of total rms amplitude in the frequency range 0.5-9 Hz. The amplitude spectra used to compose the spectrograms have been computed over consecutive, 8192 samples long, windows (sampling rate is 100 Hz), with 80% overlap. Vertical lines that appear locally at some stations may be earthquakes that were not identified when removing transients from the signal.

We computed amplitude spectra of the tremor signal over one-hour intervals. The spectra are characterised by a wide range of frequencies between 0.8 and 10 Hz. A number of peaks can be identified in the spectra, not representing overtones of a fundamental frequency, as is characteristic of harmonic tremor. We selected three frequency bands (Fig. 10) for further analysis of tremor location and amplitude:

- 0.8-1.5 Hz: This frequency range dominates the amplitude spectra in the beginning and ending hours of the tremor;

- 1.5-4 Hz: Dominant when the tremor is strongest and short bursts of tremor are recorded;
- 4-9 Hz: Lower in amplitude and not seen at some stations (e.g. ENT, ESK, SLY).

Although the overall pattern is similar at all stations, there are some exceptions. Station ENT has significantly less high-frequency content (4-9 Hz) proportionately compared to most other stations. Station ALF has a relatively even distribution of amplitude over all frequencies (0.8-10 Hz) with peaks in the amplitude spectra that can be recognized as horizontal bands in the spectrogram of Fig. 9. These features are in general similar to what is observed for the seismic events: Higher frequencies appear to be attenuated at the caldera stations, in particular at ENT, compared to station ALF.

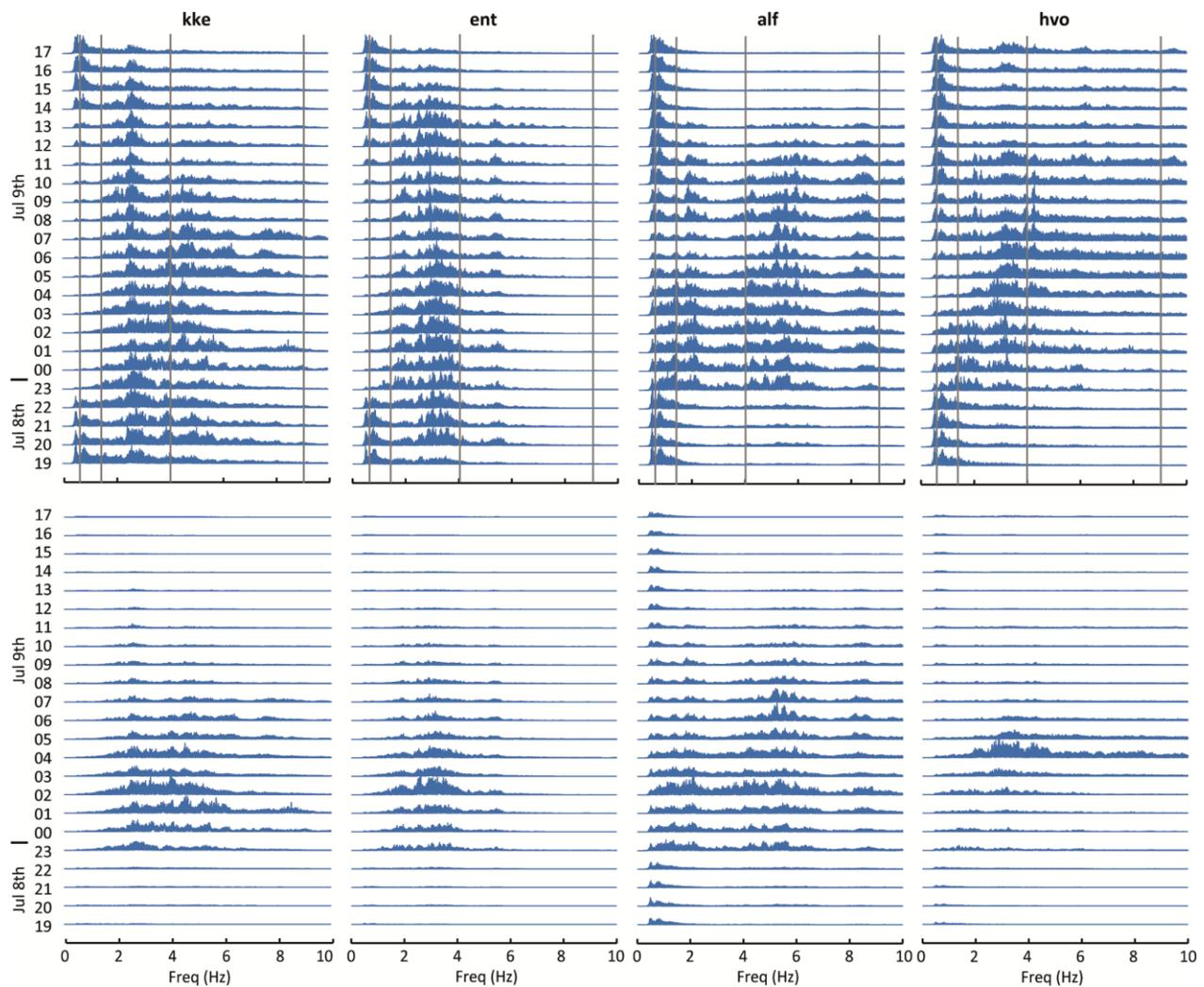

Fig. 10 Hourly amplitude spectra at some stations. Top: hourly traces individually normalised. Grey lines separate the three frequency bands. Bottom: hourly traces globally normalised for each station (all 23 hours).

The globally normalised spectra in Fig. 10 show clearly the hours when the tremor was strongest. This occurred in general between 23:00 on July 8th and 05:00 on July 9th, with the hours of tremor between 02:00 and 03:00 on July 9th dominating at all stations except for HVO. At this station, the tremor was strongest at 04:00 on July 9th, coinciding with the time of the glacial flood. Compared to the rest of the tremor, the amplitude spectrum of this different tremor phase at HVO is flatter, with power distributed more evenly over the frequency range 2 - >15 Hz. Less energy is present below 2 Hz. The width of the frequency range is greatest at the onset of this tremor phase (0.8 - >10 Hz) and then reduces gradually towards the end, around 4 hours later (Fig. 9). This coincides with decreasing amplitude as well.

## 6.2 Power time-history

In order to obtain better insight into the amplitude history of the signal at different stations, we performed a LSQ (least-squares) fit of the power time-function at each station. We aimed at identifying different components of the signal, possibly reflecting different phenomena (for example different sources).

We first computed, for each station, the integral power of the signal over 8 minute windows, sliding with 1 minute steps, between 05:00 GMT on July 8th (around 2 hours before the onset of the tremor) and 20:00 GMT on July 9th (around 2 hours after the end). We used the 2 hours before and after the tremor to evaluate the background power, which was then subtracted from the tremor power as a linear trend interpolated between the beginning and end (Fig. 11). As this led occasionally to negative power at the low amplitude stations at the beginning and end of the signal, we chose to use only 18 central hours of tremor for further analysis, as indicated in Fig. 11. As is evident from Fig. 11, the time history of the tremor has very similar features at all the stations, while the overall amplitude is quite varied. We have fitted a simple model to the tremor in order to enhance any variations.

At each station, the 3 spatial components of the 18 extracted hours have been combined into a total power (vector amplitude squared). Then a simple model was fitted to the data, parameterized as:

$$F_{ij} = a_i \cdot f_j \quad [1]$$

where $F_{ij}$ is the power time-function at station $i$, with $j$ indexing time, the function $f_j$ represents an average pattern at all stations, scaled by a station coefficient, $a_i$. This is solved through a LSQ fit which is performed in 10 minute intervals throughout the 18 hours. This fitting involves an ambiguity where an arbitrary scaling factor can scale $f_j$ relative to all the $a_i$. Thus, the method extracts relative amplitude information between stations. It is clear from Fig. 11 that most of the relative variation in amplitude occurs at station KKE. So we used this station as reference and normalised all stations' coefficients by the KKE coefficients (Fig. 12b). KKE coefficients were instead normalised by the mean of the other stations' coefficients, in each time interval (Fig. 12a). We repeated this process for each of the three different frequency bands mentioned above: 0.8-1.5 Hz, 1.5-4 Hz and 4-9 Hz.

All results, for all stations and all frequency bands, are reported in the Supplementary Material and Figs. 11-12 show some examples. Some general features are:

- The onset of the tremor is clear at about 19:15 on July 8[th] and the end is gradual, with the tremor disappearing into background noise at 18:00 the following day (Fig. 11).
- Overall, the amplitude time-history is similar at all stations, except for a scaling factor (Fig 11), and the coefficient plots are stable for most of the duration of the tremor, except for occasional peaks (Fig. 12).
- 3 main strong tremor bursts, lasting 6 to 10 minutes, occurred at around 00:30, 01.20 and 03:00 and are observed at all stations. They are dominated in terms of power at station KKE in all frequency bands, as shown clearly by the KKE coefficients (Fig. 12a).
- Another longer (40 min) burst occurred at around 01:55, recorded by all stations. This is not dominated by a particular station, as there is no corresponding peak in any of the relative coefficients plots (Fig. 12 and Supplementary Material).
- Station ENT has much less power than the other stations in the highest frequency band compared to the lower two bands (Supplementary Material).

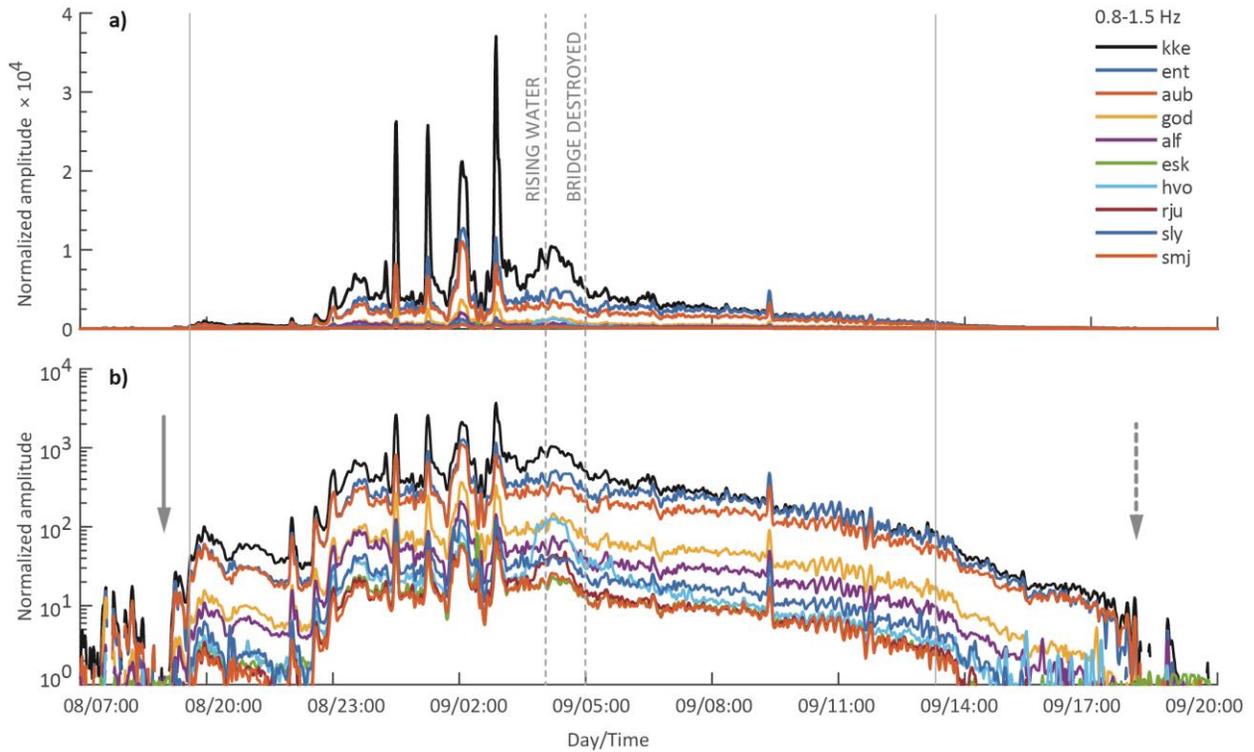

Fig. 11. a) Integral amplitude spectral density over the low-frequency band (0.8-1.5 Hz) at all stations as a function of time. b) The same plotted on a logarithmic scale. In both cases the background power has been subtracted and the amplitude normalised by the average background at station KKE. The solid arrow indicates the onset of the tremor. The dashed arrow indicates approximately the time when tremor disappears into noise. The vertical dashed lines indicate the time when rising water was observed at Léreftshöfuð gauging station (04:00) and the time when the bridge over Múlakvísl river was destroyed (05:00). Around 03:30 an anomalous tremor phase appears at station HVO. The solid lines indicate the time interval used for the LSQ fit (same as Fig. 12).

- At 03.30, a different tremor phase dominates clearly at HVO (Fig. 12b) in all frequency bands (especially the highest) and is vaguely seen at RJU in the lowest frequency range (Supplementary Material). The shape of the burst is different compared to the other short bursts described before, with a sharp increase in amplitude at the beginning, and gradual decrease in the following hours, as opposed to a sharp onset and a sharp end of the others (Figs. 11 and 12). In the medium and high frequency range, the coefficient curve at HVO appears to have a sharp peak starting at 03:30, decaying for about 3 hours and then rising up again reaching a new peak around 12:00. This second increase in amplitude is due to the seismicity that occurred locally near HVO, mentioned above in Section 5.

In addition, there are some other minor features, e.g. some other bursts with varying amplitude ratios between stations. Most of them are related to small local earthquakes which we have not managed to remove from the data.

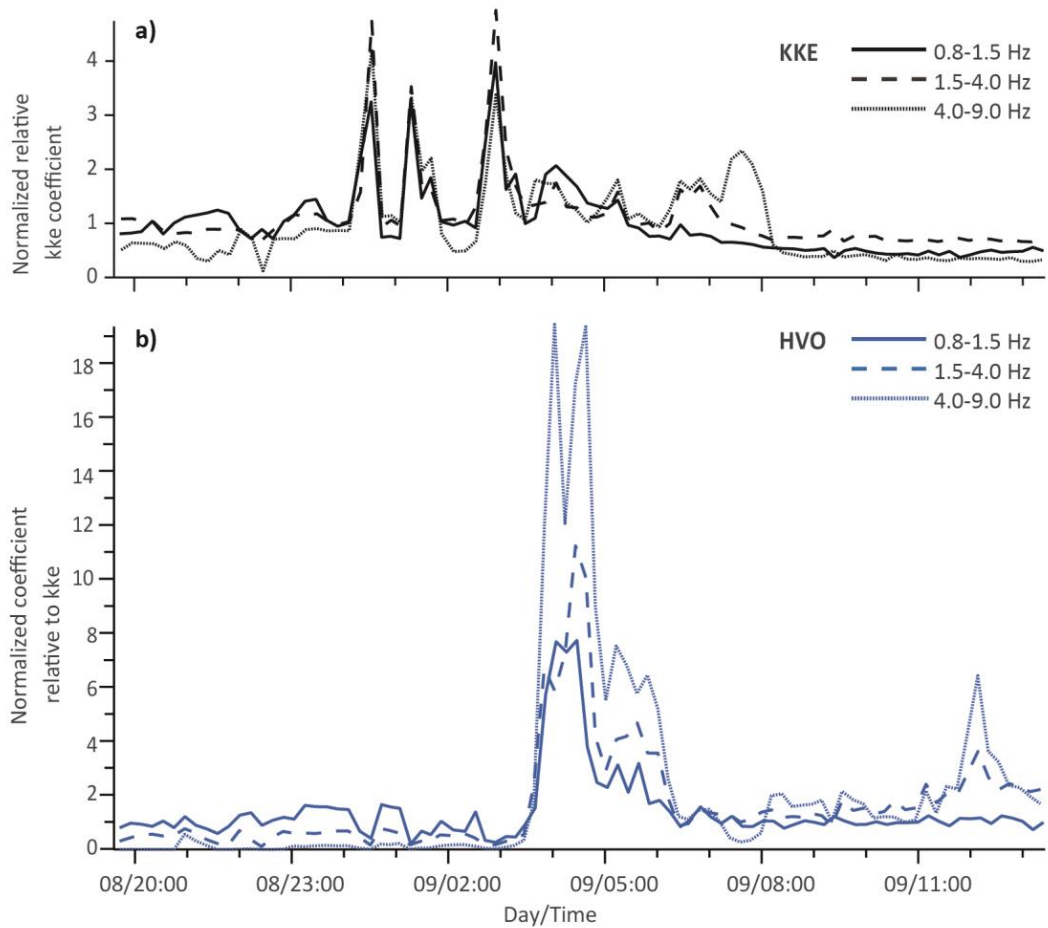

Fig. 12. Station coefficients obtained with LSQ fit of the power time-history over 10 minute intervals. a) KKE coefficients, normalised by the mean of all other stations' coefficients, in each time interval. b) HVO coefficients, normalised by KKE coefficients. Results are shown for all three frequency bands analysed.

## 6.3 Amplitude-distance decay

The amplitude decay with distance was analysed in the 3 main frequency bands identified above and by assuming a source location consistent with the southern cauldrons, most of the earthquake activity, and the main tremor source location obtained from cross-correlation, shown in the next section. The tremor amplitude was estimated as rms value of the amplitude time-history shown in Fig. 11, over the 23 hours of tremor. The small differences of station elevation are not taken into account in the distance calculation. The amplitude pattern with distance is not simple, especially at the caldera stations, and varies with frequency (blue lines in Fig. 13). In the two lower frequency bands, the relative amplitudes at stations AUB and ENT increase with distance. At higher frequency, instead, the signal amplitude at ENT drops significantly compared to all other stations and the amplitude decay with distance has a simpler pattern. This non-monotonic pattern is not consistent with both a confined location and simple distance decay.

The overall pattern is, however, stable with time and for this reason we only show the rms amplitude – distance function for the whole tremor signal (23 hours) in Fig. 13. In order to evaluate whether this pattern may be influenced by site effects, we computed the horizontal to vertical spectral ratio (H/V) at each seismic station. We did not find any clear correlation between the H/V functions and the amplitude pattern, which appears similar for all three spatial component of motion. Therefore, we present the three components combined together. Similar complexities to those of the tremor are observed also in the amplitude-distance decay of the two example events shown above in Figs. 6-7 (grey lines in Fig. 13). The peak amplitude is used as a measure of signal amplitude for these events.

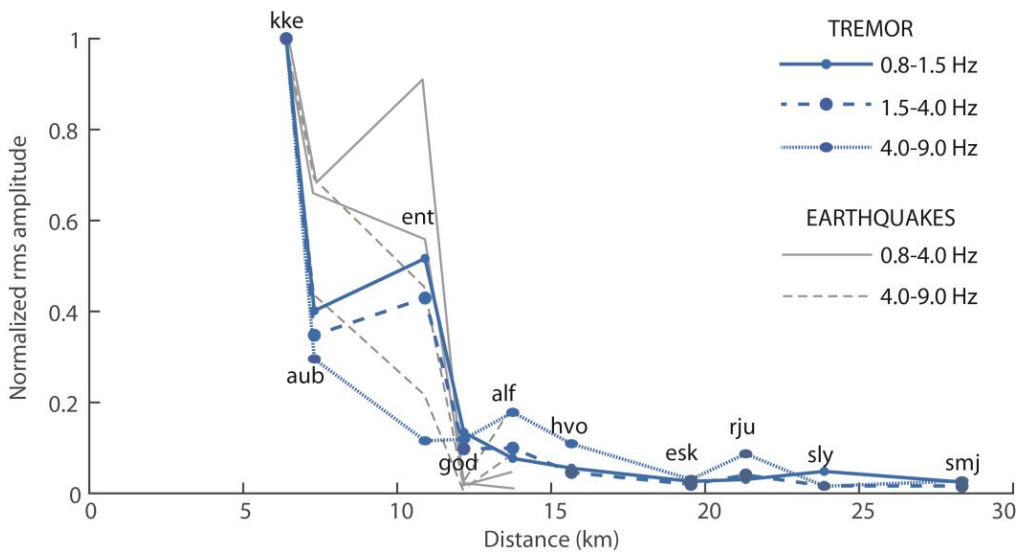

Fig.13 In blue amplitude decay with distance of the tremor for all stations in the three frequency bands, assuming the source is located at the point of highest energy in Fig. 15. Rms amplitude over 23 hours, combined for the 3 components of motion. In grey the amplitude decay with distance of the two seismic events of Figs. 6-7 for the 5 closest stations, for two frequency bands, computed as maximum amplitude combining the 3 components of motion.

## 7. Tremor source(s) location

As the amplitude pattern is not simple, especially at the caldera stations, we only used the phase information of the signals in order to locate the tremor source. Several authors have used cross-correlation analysis in order to assess the spatial distribution of ambient noise and volcanic or geothermal tremor sources (Shapiro et al., 2006; Guðmundsson and Brandsdóttir, 2010; Ballmer et al., 2013; Droznin et al., 2015). These methods perform a back projection of inter-station cross-correlations of noise/tremor records to hypothetical source locations in a geographic grid. This is done in 2 dimensions, assuming the source is located at the surface and that the tremor has a strong surface-wave component, propagating at constant velocity. We applied a similar approach, but using double instead of single correlations, as proposed by Li et al. (submitted for publication). Assuming an average uniform velocity, the double correlation of tremor recordings of triplets of seismograms (rather than pairs used in the single correlation approach) are back projected to a 2D grid of points. The results from all station triplets are then stacked for each point of the grid, resulting in a map of the stacked, back-projected correlations. This can be taken as a proxy for the energy distribution of the source, but is of course affected by the frequency and its band width, the velocity and its variation, other signals in the tremor than those that propagate horizontally, including noise, and an unknown or arbitrary amplitude scaling with distance. Li et al. (submitted for publication) addressed these issues by synthetic testing. Several different velocities have been tested and the one that best focuses the energy was chosen. This velocity (1.2 km/s) is slightly lower than Rayleigh-wave group velocities measured at 0.5-1 Hz at Hekla (Haney et al., 2011) and at Katla and nearby Eyjafjallajökull (Z. Jeddi and Á. Benediktsdóttir, personal communication)

In order to reduce the effect of transients in the signal, we processed the tremor using a 1 bit normalization (Bensen et al., 2007; Li et al., submitted for publication) and this proved to work better than clipping or manually removing earthquakes from the tremorgrams. In fact, this focuses the energy of the source more effectively, reducing the influence of the nearby stations, which otherwise dominate in terms of amplitude with respect to all other stations.

We applied this method to the lowest frequency band (0.8-1.5) because at higher frequency it is more difficult to obtain stable results and the low frequencies are less likely to be affected by scattering effects.

## 7.1 Cross-correlation functions

The cross-correlation functions for all station pairs have been computed over one hour records (Fig. 14). The pattern of correlation is complex. If the sources were truly diffuse, and the wavefield dominated by surface waves, we would expect two symmetric wave packets at opposite time shifts, corresponding to intra-station surface waves. This is not the case. If the source area was geographically small and there was no multipathing, we would expect to see an isolated wave packet at a time corresponding to the difference in distance of the two stations from the source divided by an average wave velocity. Instead, several wave packages can be identified in the cross-correlation functions, distributed over a wide range of time shifts, not symmetric. The correlation functions are also not stable with time, in particular between 11:00 on July 8th and 04:00 on July 9th, when the tremor is strongest. During this period, a number of wave packages can be identified in all cross-correlation functions distributed over time shifts of several tens of seconds. After this unstable period, the functions become much more stable until the end of the tremor, with few wave packages dispersed over 10-15 seconds. Auto-correlations of tremor at individual stations do not suggest significant source correlation on these time scales. The wide time dispersal cannot be explained by direct propagating waves from a single point source only. Higher-order scattering effects must be invoked. Multiple wave packages at smaller time shift may be caused by a distributed source or multiple sources.

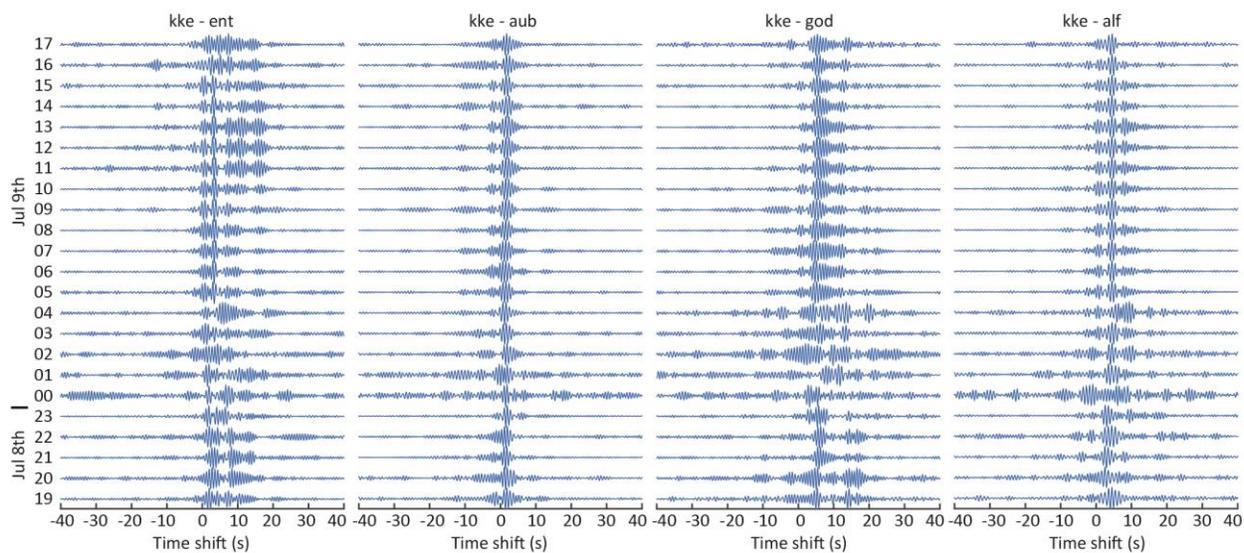

Fig.14 Examples of hourly cross-correlation functions for four station pairs. Several wave packages can be identified, dispersed over wide ranges of time shift. Temporal variations of the tremor can also be noticed. An unstable period between 00:00 and 05:00 is followed by a more stable period until the end of the tremor.

### 7.4 Tremor location results

The source distribution of energy obtained with the double-correlation method by Li et al. (submitted for publication) for the whole tremor signal is shown in Fig. 15a. We tested a range of velocities between 1 and 2.5 km/s and chose 1.2 km/s as the velocity that best focused the energy. The energy peak in the south-eastern caldera corresponds to the inferred location of the tremor source. This is consistent with the location of the earthquakes that occurred during the tremor and with the locations of the southern active ice cauldrons (Fig. 15a). The width of the maximum energy peak is affected by several factors: The finite width of the frequency band used, the uncertainty of velocity, the size of the source. A part of the energy is noticeably distributed along the hyperbolae corresponding to constant time shifts for the 3 caldera stations' (KKE, ENT and AUB) pairs. This is expected, as they are the three closest stations to the source and dominate not only in terms of signal amplitude, but also in terms of coherency.

As there are three clear short tremor bursts that showed a distinct behaviour in the relative amplitude, we isolated them and located them separately from the rest of the tremor. We used the same double-correlation method, applied to each 6-min burst. We then stacked the three energy-maps to better suppress noise. By doing this, we assume that the three bursts were generated at the same location. This is justified by the relative amplitude behaviour, where all three peaks have the same pattern, different from the average pattern at other times, suggesting a similar source location. The result is shown in Fig. 15b: although the energy is not as focused as in the case of the whole 23 hour signal, a peak of energy is located to the north-east of the main tremor source located in the south-eastern caldera. This correlates with the north-eastern active cauldron and with the evidence described in the previous section that the power related to these peaks is dominated by station KKE, which is located very close to this cauldron.

It was not possible to locate the tremor component generated by the flood, for several reasons: i) its presumed location is peripheral to the network, ii) it is observed mainly at one station and weakly at a few other stations, while at most stations it is hidden in the main tremor , iii) the source may not be stable in space.

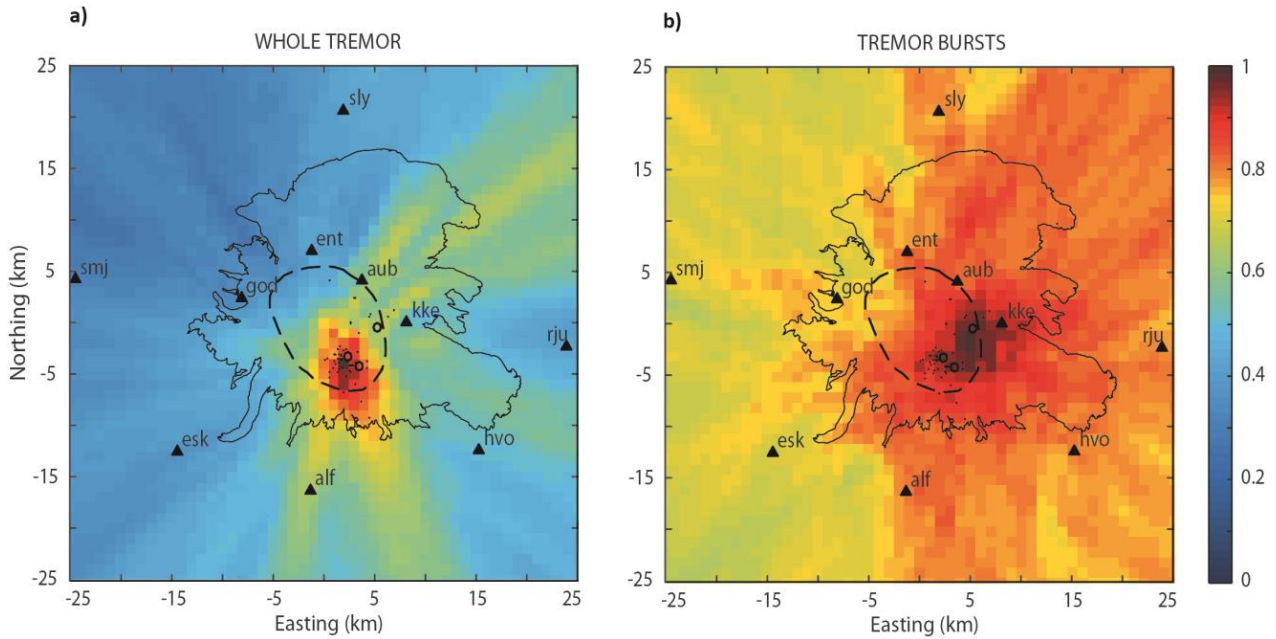

Fig. 15. Location results with double-correlation method, obtained for a) the whole tremor signal and b) only the short tremor bursts, with velocity 1.2 km/s. The colours define normalised energy, dark red for maximum energy. Black circles: ice cauldrons that collapsed during the tremor. Black dots: earthquakes recorded during the tremor. Black solid line: glacier. Black dashed line: caldera outline. Black triangles: seismic stations.

## 8. Discussion

### 8.1 Volcanic/hydrothermal tremor and flood tremor components

Most of the tremor signal appears to be generated at a stable source located in the south-eastern part of the caldera, consistent with the two southern active cauldrons and with the earthquake activity. This component of the tremor likely corresponds to the stable portion of the relative amplitudes at various stations (Fig. 13 and Supplementary Material). This suggests that there is a main source of tremor which was stable in space and time, although varying in frequency content and power, probably in association with changes in the source mechanism. This source might have been spatially distributed, based on the width of the peak of energy obtained from back-projection of the cross-correlation functions (Fig.15). However, part of this width is due to uncertainty and a part due to the finite width of the frequency band used. We are not able to estimate the uncertainty rigorously, but the average half width of the distribution in Fig. 15a is about 3 km.

In addition to this main source, another component of the tremor was identified with the LSQ fit of the tremor power, corresponding to the three short tremor bursts, dominated in amplitude by station KKE (Fig. 13a). This component has a similar amplitude pattern in all frequency bands analysed. The location results obtained with back-projection of the double-correlation envelopes highlighted a peak of energy correlating in space with the northern active ice cauldron, located close to station KKE (Fig. 15b). We therefore suggest that another source of tremor, located closer to KKE, was either intermittently activated, or intermittently exceeded the main source's power.

A third distinct tremor phase is primarily seen at station HVO (Fig. 13b), the station closest to Múlakvísl river. This phase correlates in time and space with the water flood and therefore appears to be directly related to the water draining from Kötlujökull glacier.

The evidence of a stable tremor phase, both in time and space, located near active cauldrons and clearly separated from a tremor phase associated with the flood, suggests that most of the tremor was generated by volcanic or hydrothermal processes occurring at the location of the active cauldrons. This is also supported by the increased earthquake activity in the same location and by the water accumulation under the glacier that started months before the tremor, which is explained by increased geothermal activity (Guðmundsson and Sólnes, 2013). In addition, the power of volcanic tremor is often concentrated in the band between 0.5-7 Hz (Konstantinou and Schlindwein, 2002), which is consistent with our observations. The same applies to the source located in the eastern sector of the caldera, corresponding to the short tremor bursts. Here the source appears less stable in terms of amplitude history, but stable in space, allowing us to crudely estimate a location that corresponds to the eastern active cauldron.

The two different types of tremor source suggested (flood-related and volcano-related) may be reflected also in different amplitude and frequency features of the signal, summarised here:

- the flood tremor spans a wider frequency range, with energy up to >15 Hz, while the volcanic tremor has energy up to 9-10 Hz;
- the frequency content of the flood tremor is flatter, with most of the energy evenly distributed over a wide range (2-10 Hz), while the hydrothermal/volcanic tremor is mainly concentrated between 0.8 and 4 Hz;
- the flood tremor has the widest frequency range in the beginning and gradually loses high frequencies as the signal decays in amplitude, while the volcanic tremor has a more stable frequency distribution through time, which does not correlate with changing signal amplitude;
- the flood tremor begins with large amplitude which monotonically decays with time over a few hours, while the volcanic tremor has a more complex amplitude history.

## 8.2 Interpretation of the volcanic/hydrothermal source

Possible interpretations of the source generating the tremor located at the active ice cauldrons are either a subglacial magmatic eruption or a hydrothermal process, such as hydrothermal boiling, eventually involving explosive events. We have insufficient evidence to distinguish whether or not a minor subglacial eruption occurred. Certainly, an increase in geothermal heat release has occurred, starting about one year before the tremor episode (when water accumulation started under the glacier) and a subglacial eruption is not an unlikely scenario, considering that the event was also accompanied by greatly increased seismicity inside the caldera.

The 2011 unrest was similar to the event that occurred in 1999, that some authors interpreted as a subglacial eruption (e.g. Guðmundsson et al., 2007). For the 1999 event, the arguments to support the hypothesis of an eruption are that the heat exchange that led to the formation of the melt water occurred very rapidly (few hours or days) and that there was no appreciable geothermal heating at the same site in the following years (Guðmundsson et al., 2007). This is different from what happened in July 2011, as in this case geothermal activity was observed for years before 2011 and still persists. However, this does not exclude the possibility that, if in 1999 a subglacial eruption took place, a similar episode occurred in 2011. Although in both cases increased seismicity was observed, neither of the two unrest episodes showed clear seismic indications of magma rising. In addition, geochemical studies of the flood water from the 2011 jökulhlaup did not find evidence for floodwater having come into contact with magma (Galeczka et al., 2014).

Another possible scenario is that the tremor was generated by hydrothermal processes. For example, hydrothermal boiling generating tremor may have been induced by the pressure drop that occurred when the water level in the subglacial lakes dropped as a consequence of water release from the cauldrons. The tremor started at 19:00 on July 8$^{th}$ and the flood waters reached the gauging station at Léreftshöfuð 9 hours later, at 04:00 on July 9th. It is difficult to evaluate how long it may have taken for the water to flow from the cauldrons to the gauging station, as this strongly depends on the unknown subglacial water drainage system and topography. As a reference, we used two known jökulhlaups which occurred at Eyjafjallajökull in 2010 and Gjálp in 1996. The first occurred during the 2010 Eyjafjallajökull eruption and took about 5 hours to reach a proglacial lake around 5 km away from the crater, down a steep slope (Magnússon et al., 2012). During the Gjálp eruption, instead, the flood took approximately 10 hours to travel about 50 km on a more gentle slope (Einarsson et al., 1997). The difference between these two cases may depend on the time of the year when the eruption occurred, influencing the subglacial drainage system. While the Eyjafjallajökull eruption occurred at the end of the winter, when the drainage system is inefficient, the Gjálp eruption occurred at the end of the summer when the system is fully developed (Magnússon et al., 2012). The unrest at Katla occurred in early summer and the distance between the cauldrons and the Léreftshöfuð gauging station is around 20 km. The steepness of the slope is intermediate between the two reference cases. Using the two examples as extremes, the Katla flood might have taken between 4 to 20 hours to reach the gauging station at Léreftshöfuð. This suggests that it is plausible that hydrothermal boiling, generating tremor, was initiated when water was released from the active cauldrons, reaching the first gauging station 9 hours later. This interpretation, however, does not clearly explain the increased earthquake activity that started some days before, on July 6th, inside the caldera and the new seismic cluster on the south flank (Sgattoni et al., 2016).

    The tremor signal is highly variable in amplitude. Tremor amplitude variations at other volcanoes have in many cases coincided with visual observations of varying strength of volcanic/hydrothermal activity (e.g. lava fountaining or dome building), as for example at Kilauea (Dvorak and Okamura, 1985) and Hekla (Brandsdóttir and Einarsson, 1992). However, this is not always the case and sometimes no relationship between surficial activity and amplitude has been identified. This has been interpreted as a consequence of variation of magma flow rate at depths in the crust, e.g. at Kilauea (Ferrazzini and Aki, 1992). In the case of Katla's tremor, the short tremor bursts, which seem to be located at a different site compared to the main tremor source, appear to be the strongest in terms of power and occur with a sharp onset and sharp end. If the source was hydrothermal, this might be explained with local, more powerful hydrothermal explosions or flash-boiling. The possibility to generate hydrothermal explosions depends on local conditions such as

permeability. A local low-permeability layer, for example, could induce a build-up of pressure, suddenly released into steam flashing (Morgan et al., 2009). The conditions for this to happen may have occurred only at the site of the northern tremor source, explaining the higher amplitude bursts generated there. If the tremor source was an eruption, the amplitude variations may be explained with eruption phases of varying strength.

## 8.3. Considerations about path effects and location method

The frequency-dependent, complex pattern of amplitude decay with distance, not following any clear amplitude-distance decay law, suggests that source radiation is anisotropic and/or propagation effects are complex. A similar pattern is seen also for the earthquakes. Site effects may play an additional role, but no clear correlation was identified between H/V spectral ratios and the amplitude distribution at the different stations. Also, the amplitude pattern is similar for all components of motion, which may indicate that site effects do not play the main role. However, the H/V spectral ratio method is usually used to detect amplification due to resonance in a stratified structure. The caldera stations were all deployed on nunataks in the ice, on sharp bedrock peaks protruding the ice sheet. The elastic properties of ice are rock-like, but its density is much less than that of rock. The potential amplification effects of such topographic features are poorly understood. We, nevertheless, suggest that path effects and/or an anisotropic radiation pattern are the main responsible factors for the complex amplitude patterns observed. In addition, there is clear indication of strong propagation effects, within the caldera, suggested by the strong attenuation of high frequencies at stations receiving seismic ways travelling through the caldera region, for both tremor and earthquakes. The complexity of the cross-correlation functions is also indicative of strong scattering effects, generating several, broad wave packages in the correlation functions. The interaction of seismic waves with the subglacial topography and the ice, together with the crustal heterogeneities, may be responsible for strong path effects.

The complex amplitude pattern made the use of amplitude-based tremor location methods impossible. However, by using the signal phase through a double-correlation method, we were able to confidently locate two tremor sources (for the lowest frequency component of the tremor), in locations that are consistent with other seismic and hydrological observations.

## 9. Conclusions

We have analysed the 23 hour tremor signal and earthquake activity associated with an unrest episode that occurred at the subglacial volcano Katla in July 2011. During the unrest, three ice cauldrons deepened on the glacier and a glacial flood caused damage to infrastructure, but no visible eruption broke the ice surface. Three different tremor components were identified based on amplitude and frequency features. We have described them in detail and discussed their possible source process by using additional hydrological observations and comparison to other case studies. Back-projection of double cross-correlation functions was used to locate the two spatially stable components of the tremor.

The main conclusions of this work are:

- Increased earthquake activity, characterised by low-frequency, hybrid and high-frequency events, started inside the Katla caldera a few days before the tremor burst and lasted for months afterwards;
- The tremor signal can be separated into three main phases. Two of them were traced to the active ice cauldrons and are interpreted to be caused by hydrothermal or volcanic processes. The third, mainly observed at the station closest to the river that flooded, is associated with the glacial flood;
- Because of the highly increased seismicity, evidence of rapid melting of the glacier and similarity to the 1999 event that was interpreted as a subglacial eruption, the 2011 tremor may have been caused by a minor subglacial eruption;
- It is also plausible that the tremor was generated by hydrothermal processes with no magma involved: boiling and/or explosions may have been triggered in the hydrothermal system when the flood started to flow out of the subglacial geothermal systems;
- All interpretations require an increase of heat released by the volcano that led to water accumulation before the tremor. This may be due to heat introduced into the shallow crust by a magmatic process or enhanced permeability in the geothermal areas due to tectonic activity;
- The complex amplitude-decay with distance precluded the use of amplitude information to locate the tremor sources and suggests the presence of strong path effects on waves travelling through the caldera region. This is corroborated by strong attenuation of high frequencies along trans-caldera paths.

## Acknowledgements

The authors would like to thank the Icelandic Meteorological Office for access to waveform data. The temporary deployments producing data for this study were supported by CNDS (Centre for Natural Disaster Science, www.cnds.se) at Uppsala University and the Volcano Anatomy project, financed by the Icelandic Science Foundation. This work was funded by the University of Bologna, University of Iceland and Uppsala University, as a part of a joint PhD project.

## Supplementary material: Appendix A

Figs. A1, A2, A3. a) Integral amplitude spectral density at all stations as a function of time. b) The same plotted on a logarithmic scale. In both cases the background power has been subtracted and the amplitude normalised by the average background at station KKE. c) and d) station coefficients obtained with LSQ fit of the power time-history over 10 minute intervals. c) KKE coefficients, normalised by the mean of all other stations' coefficients, in each time interval. d) all other stations' coefficients, normalised by KKE coefficients. Results are shown for the three frequency bands: 0.8-1.5 Hz in Fig. A1, 1.5-4.0 Hz in Fig. A2, 4.0-9.0 Hz in Fig. A3.

**Figure A1**

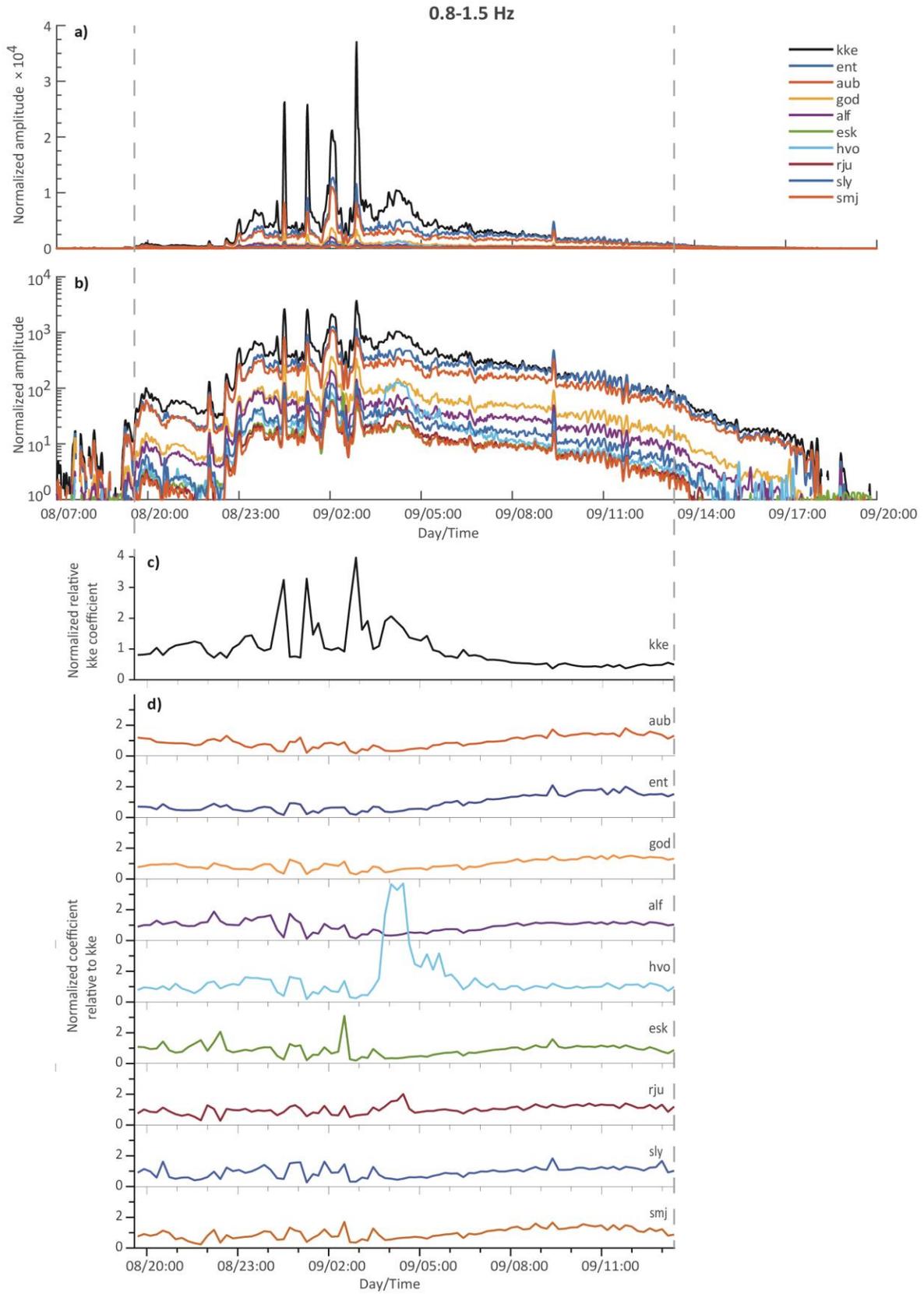

**Figure A2**

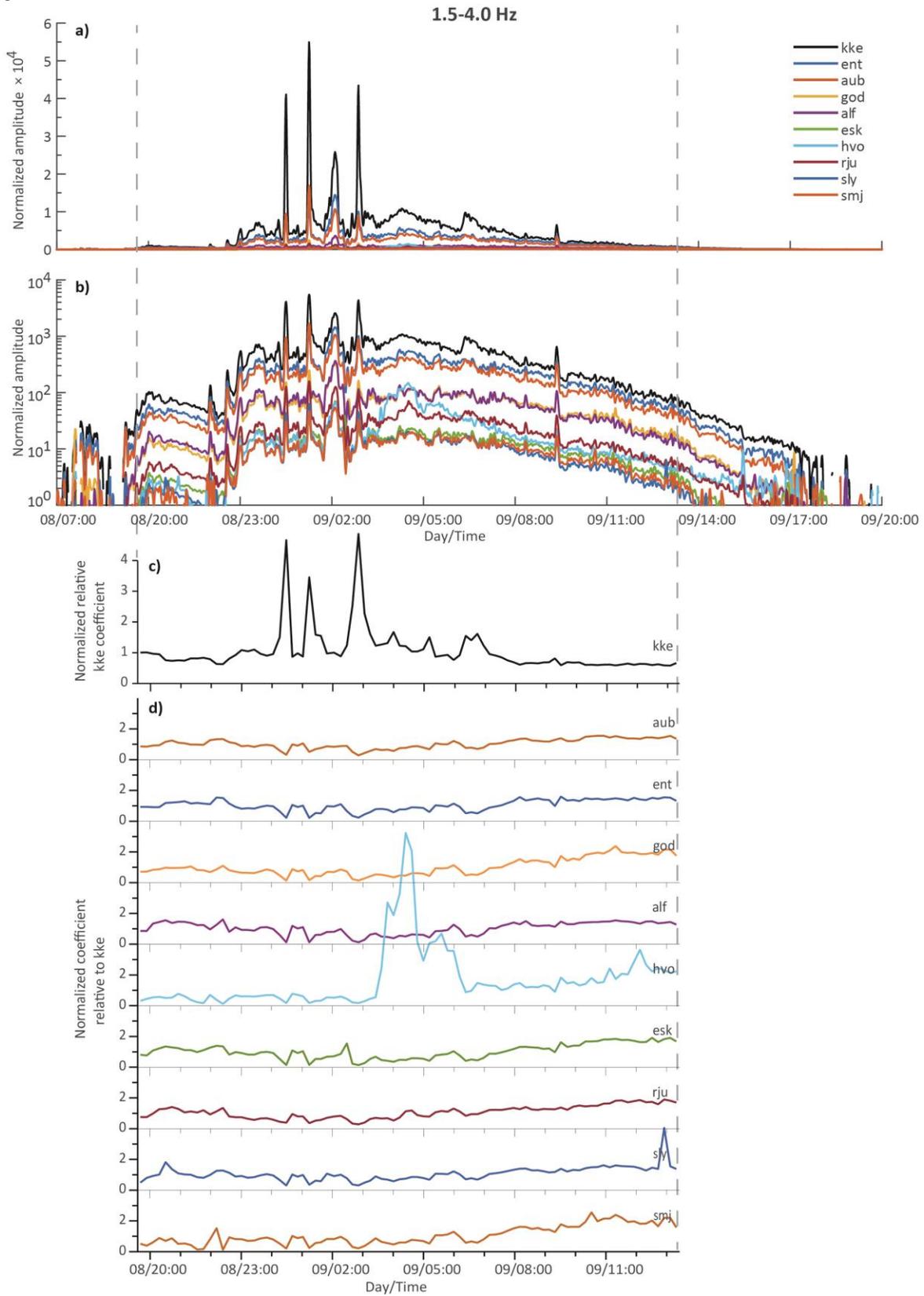

**Figure A3**

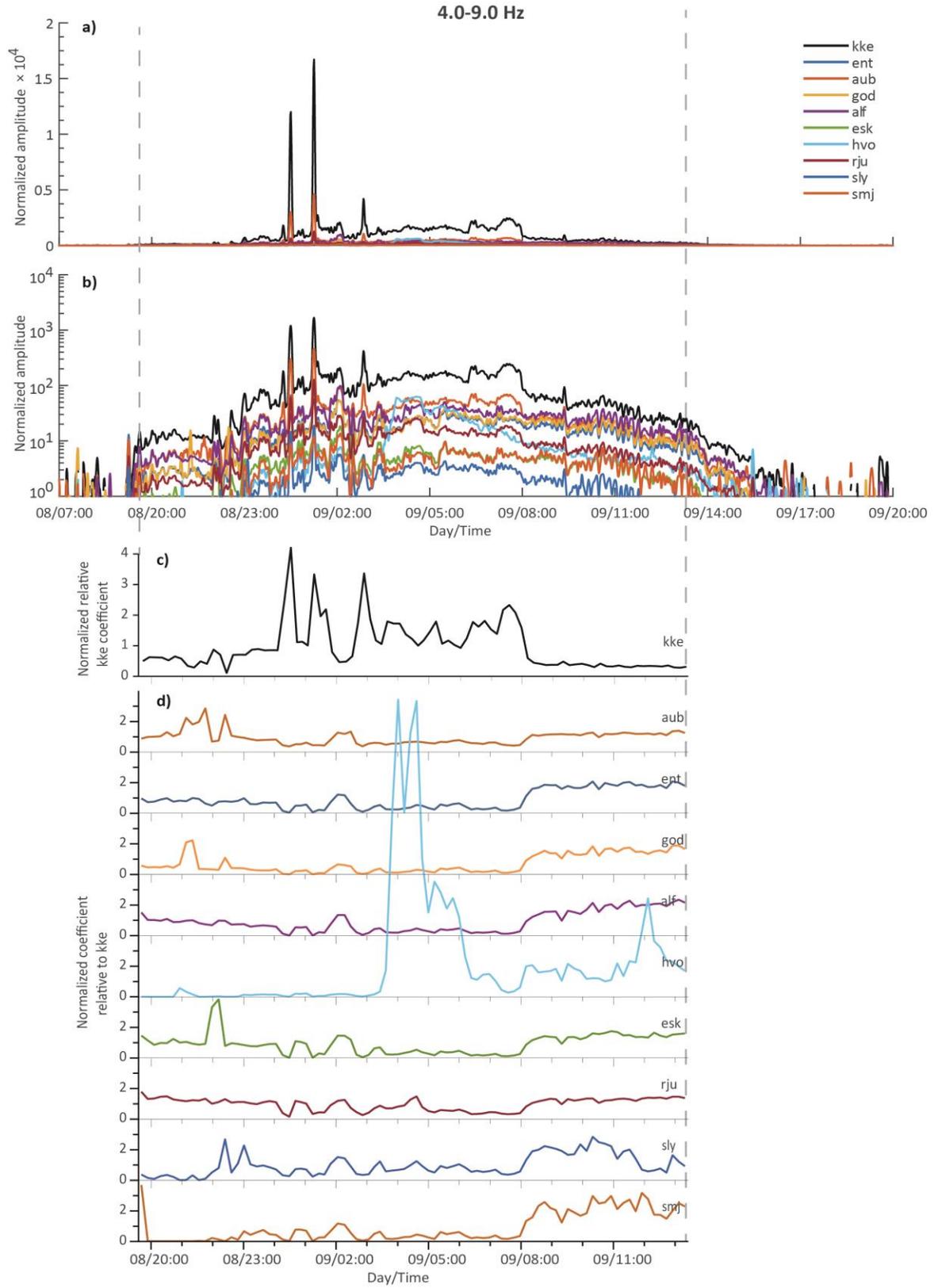